\documentstyle[11pt,newpasp,twoside, epsf]{article}
\def\edcomment#1{\iffalse\marginpar{\raggedright\sl#1\/}\else\relax\fi}
\reversemarginpar

\begin{document}
\title{Patterns of Super Star Cluster Formation in `Clumpy' Starburst Galaxies}
 \author{J.S. Gallagher, N.L. Homeier, C.J. Conselice}
\affil{Department of Astronomy, University of Wisconsin-Madison, 475 N. Charter
St. Madison WI., 53706}
\author{WFPC-2 Investigation Definition Team}

\begin{abstract}

We present preliminary results from a
Hubble Space Telescope (HST) WFPC2 investigation of spatial and
temporal distributions of star clusters in the 
clumpy irregular galaxy NGC~7673 and the starburst spirals NGC 3310 and
 Haro~1.   We compare
the spectral energy distributions of star clusters in the large
clumps in NGC 7673 to model calculations of
stellar clusters of various ages. We also propose that the presence of 
super star clusters in clumps seems to be a feature of intense starbursts.

\end{abstract}


\section{Introduction}

    The optical morphologies of active star forming galaxies are often
dominated by kpc-scale star-forming regions, which appear as luminous
`clumps' or `islands'. These clumps can have very high optical and UV
surface brightnesses, making them visible signatures of intense star
formation over cosmological distances.  High angular resolution imaging
with the {\it Hubble Space Telescope} (HST) and ground-based telescopes,
reveals that clumps consist of associations of dense star clusters.
These are usually superimposed on a more diffuse background, probably
made up of massive stars. 

The presence of numerous  star clusters influences the evolution of 
clumps through their
interactions with the surrounding ISM, that must respond to
photoionization, as well as mechanical energy and momentum inputs from
the evolving star clusters. These processes can trigger star formation
and simultaneously remove ISM from the clump. Since star clusters can
be age-dated from their spectra, their presence also offers the
possibility of measuring the history of
star-forming activity in starburst galaxies.

Starbursts are well-known producers of dense, massive star clusters and thus
are an excellence place to study these interactions. In 
this paper we explore the spatial and temporal patterns of massive 
star cluster formation in a small sample of relatively nearby starburst 
galaxies.

The primary objects in this study are NGC 2415 (Haro 1), NGC 3310 and
NGC 7673. The Haro 1 and NGC 7673 data are HST WFPC2 ultraviolet (UV)
and optical images, while we use
ground-based optical images from the WIYN 3.5-m telescope for NGC 3310.
Details are presented in Gallagher et al. (2000).

\begin{table}
\begin{center}
Properties of Studied Starbursts
\begin{tabular}{rlll} \hline
Galaxy & Distance & Magnitude & Size   \\ \hline \hline 
NGC 2415    & D=50 Mpc &  M$_B$=-20.8 &  A=13 kpc \\

NGC 3310    &       D=19 Mpc & M$_B$=-19.9 & A=17 kpc \\

NGC 7673    &       D=46 Mpc &  M$_B$=-20  &  A=17 kpc \\
\hline
\end{tabular}
\end{center}
\end{table}

\section{Star Clusters in Starbursts}

We begin with NGC 3310, the least intense example in our sample (Figure 1). 
WIYN optical images reveal compact star clusters in the ring and along the 
arms of this galaxy, 
with a wider distribution of fainter cluster candidates outside of these
regions. While clusters are numerous, their spatial distribution in this 
non-clumpy starburst resembles that of a normal spiral galaxy. 

Haro 1 is an intense starburst with high optical brightness, producing 
a luminous galaxy with a diameter of only about 13 kpc. The spectrum shows
strong Balmer absorption lines indicating that this is a relatively 
evolved starburst. High quality ground-based images with the WIYN
Telescope suggest a transition to a clumpy structure. Our WFPC2 image
in the emission line-free F547M filter displays a moderate
degree of clustering of dense, luminous star clusters, and high surface 
brightness regions usually containing more than one compact star cluster.
{\small
\begin{figure}
\plottwo{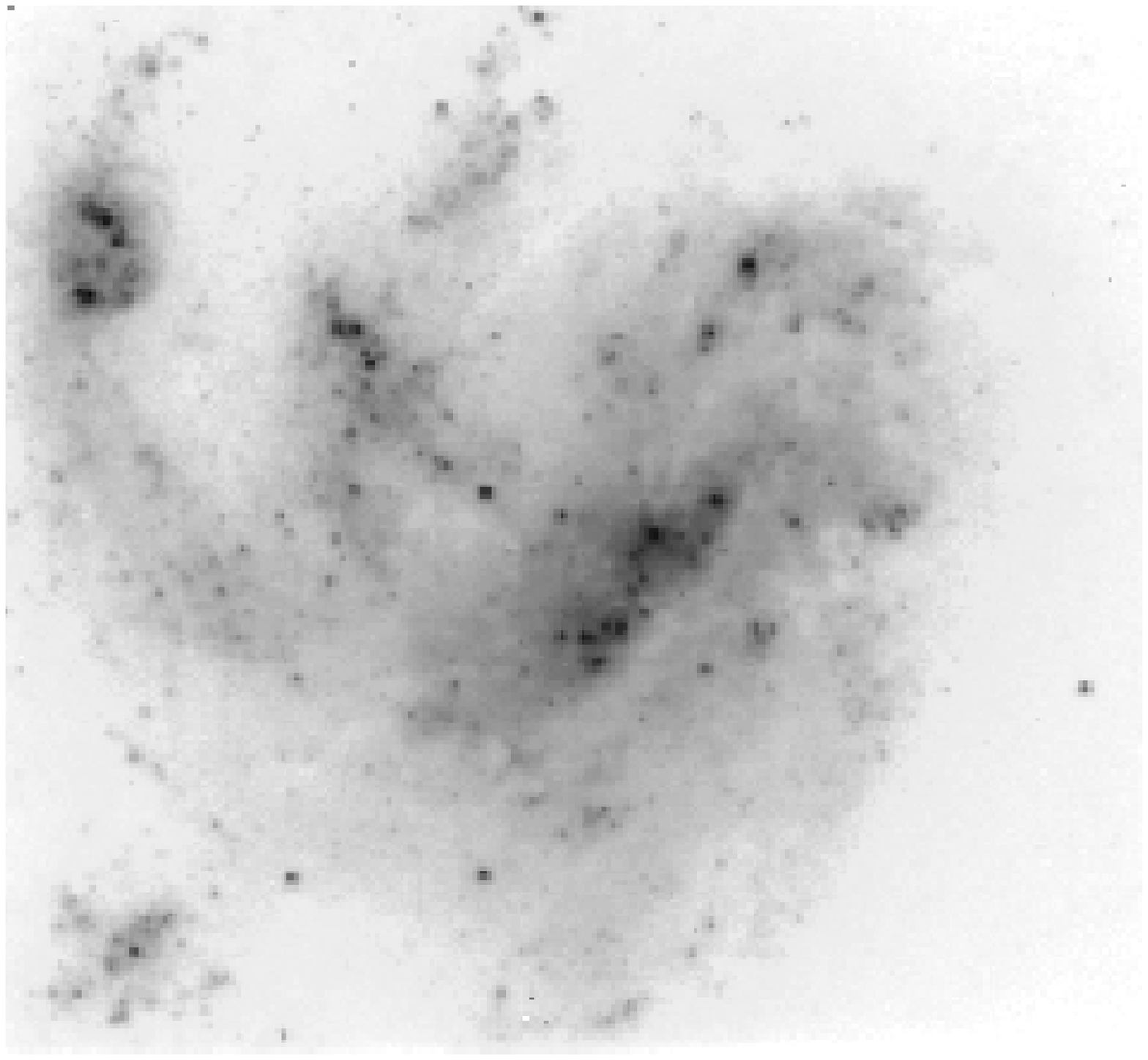}{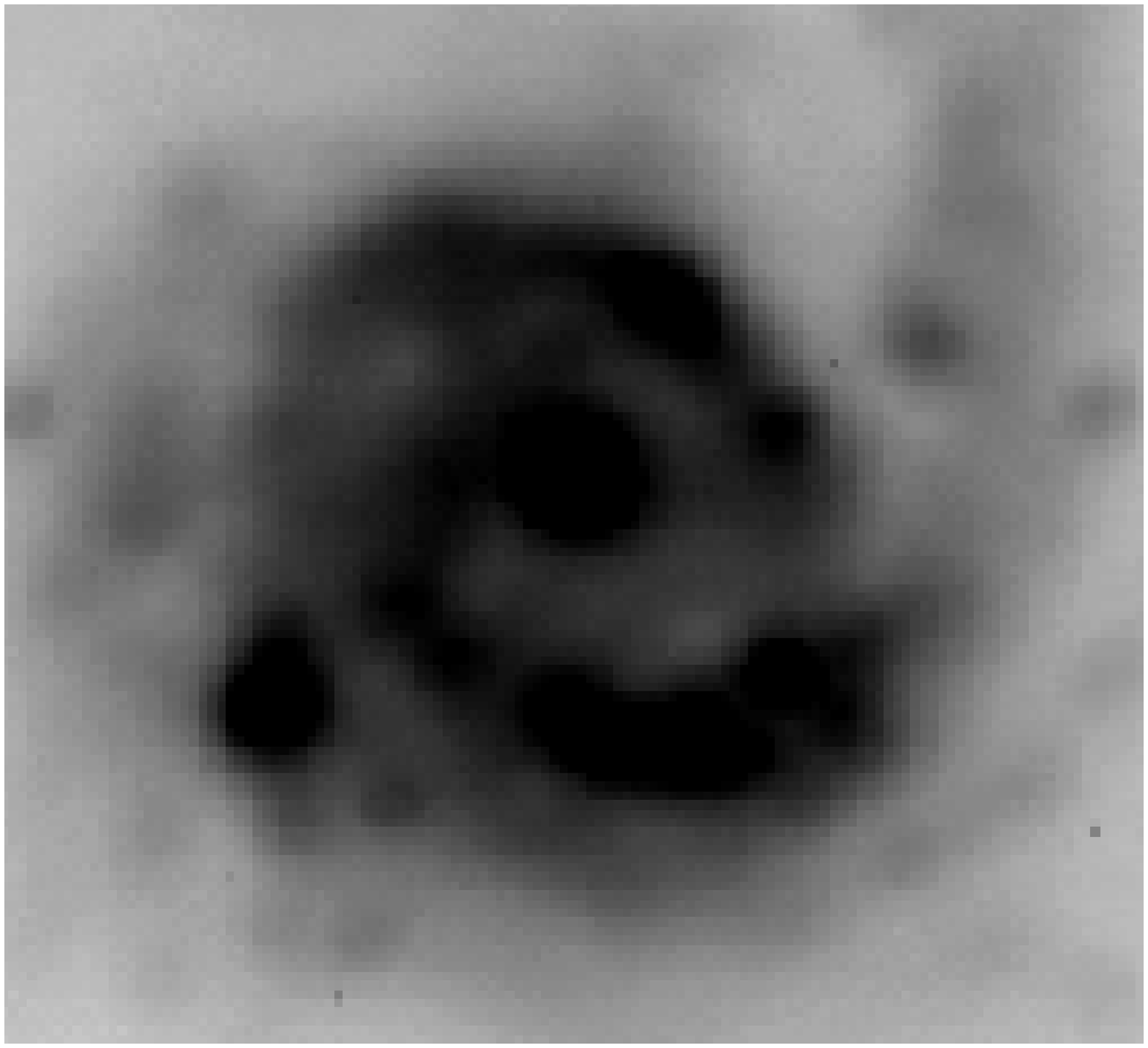}
\caption{WFPC2 image of NGC 7673 (left) showing the clumpy star formation
appearance that is occurring in distinct star clusters. A WIYN image
of the center of NGC 3310 (right) displays a more normal circumnuclear ring 
of star formation.}
\end{figure}
}

\begin{figure}
\plottwo{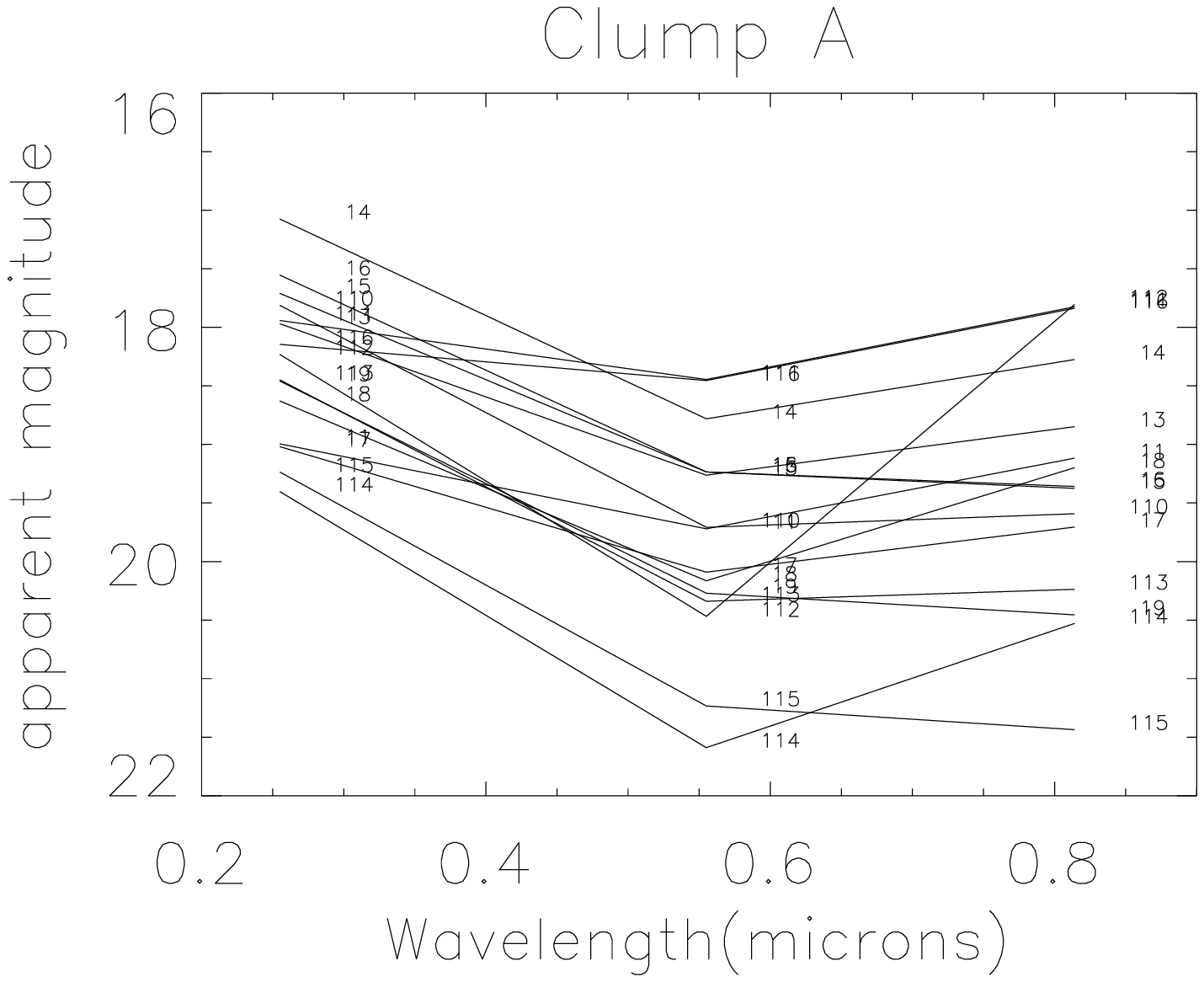}{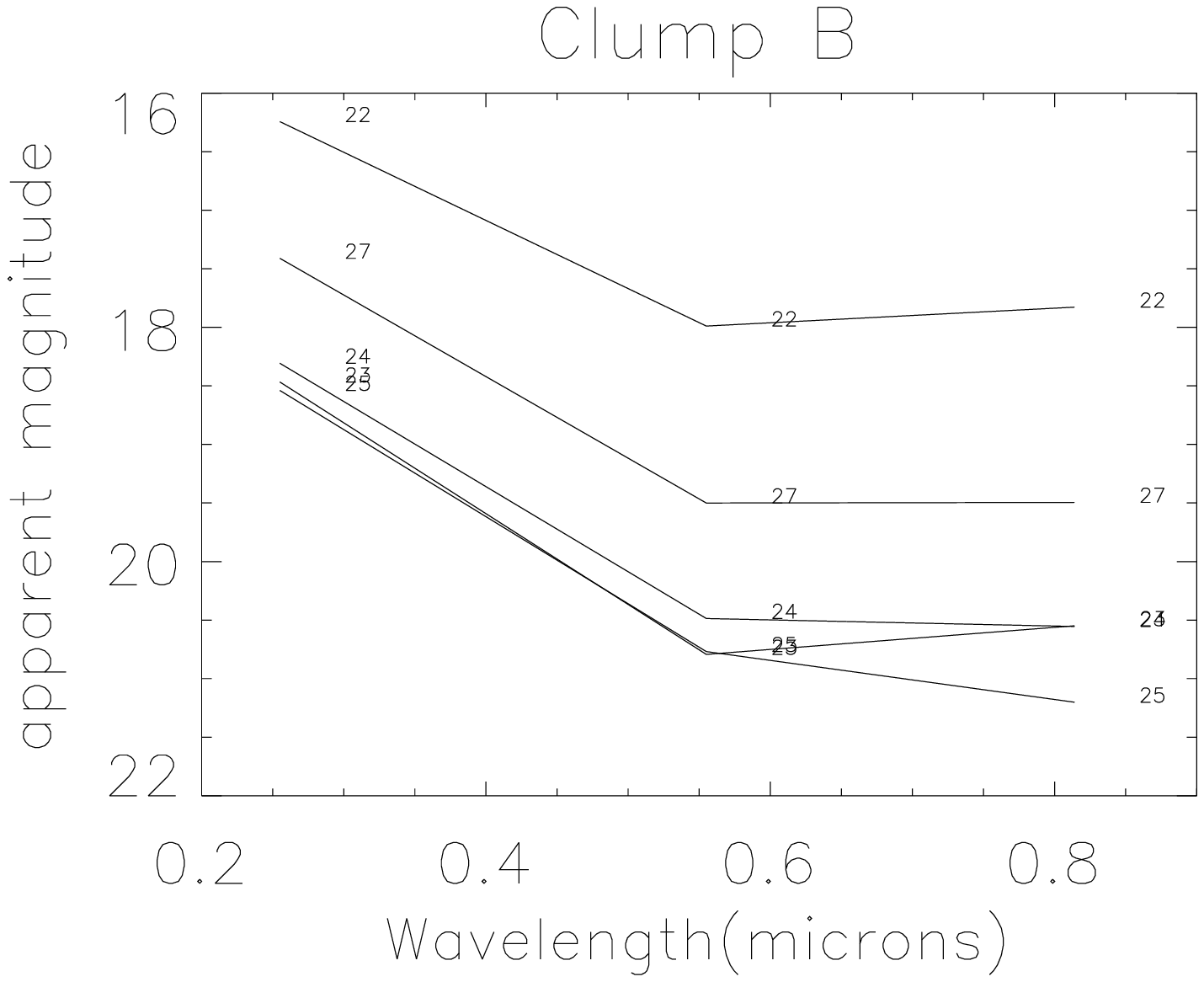}
\caption{Spectral Energy distributions for super star clusters
within two clumps in NGC 7673.}
\end{figure}

\begin{figure}
\plotone{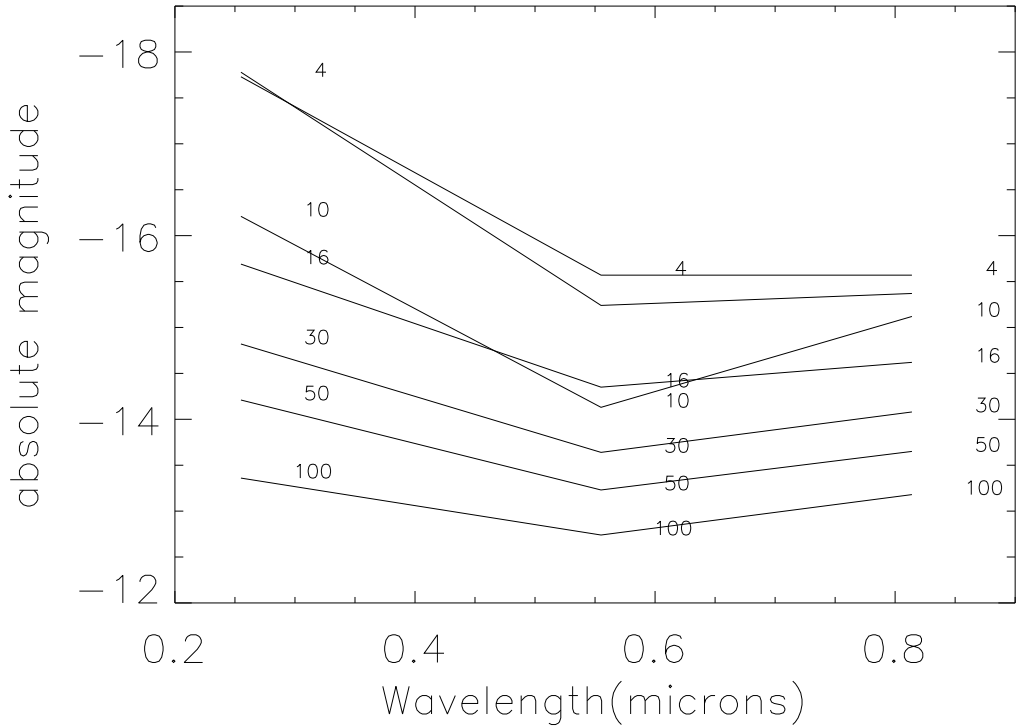}
\vspace{1cm}
\caption{Model calculation for an aging cluster from 4 Myr to 100 Myr (right
y-axis) vs. wavelength. Fluxes from Starburst99, and magnitudes from
models by A. Watson.}
\end{figure}

NGC 7673 is a star bursting clumpy irregular galaxy.
 Colors and magnitudes derived for star
cluster candidates in the F255W, F555W, and F814W WFPC2 bands indicate
that young clusters are mixed through the optically prominent clumps; 
their colors appear to be
driven at least as much by locally variable extinction as by ages. A low
level of organization may be present, with clusters near the galaxy's center
possibly being made in a linear structure by bar-induced gas flows. Some
clumps are roughly circular, and so propagating star formation could be 
present.  Figure 2 shows the spectral energy distribution (SED) of two
clumps in NGC 7673, while Figure 3 shows model calculation SEDs
for star clusters at various ages. 
Probably more than one mechanism structures the distribution of star
clusters within clumps, but in most cases populations of star clusters 
are formed over relatively short time scales of $<$ 100 Myr.

\section{Discussion}

Clumps seem to be features of active star bursts with high intensities.
This fits with theoretical models (Elmegreen \& Efrenov 1997; Noguchi
1999) where stellar clumps form in gas-rich galactic disks
with higher than average internal velocity dispersions, leading to large
Jeans masses and lengths. Such conditions could be natural consequences of
collisional perturbations of gassy disk galaxies. They would be more
common in less evolved galaxies with higher gas contents and lower degrees
of organization, providing a possible explanation for the clumpy
appearances of high redshift galaxies (Noguchi 1997).
Although super star clusters can form in a range of conditions, clumps may
be particularly important in unevolved galaxies; observations of nearby
clumps then may provide insight into the formation of globular clusters.

Once star formation begins in a clump, the subsequent evolution must be
complex. Energy and momentum inputs from massive stars and star clusters
produce obvious observational signatures in emission line profiles
(Homeier \& Gallagher 1999) will disturb the ISM, and
likely lead to compressed regions which are excellent sites for further
cluster formation (eg., Scalo \& Chappell 1999). The close
spacing between clusters and high densities of gas clouds may also lead to
cluster-cluster mergers or cluster rejuvenations due to gas cloud
captures. These possibilities reinforce the capabilities for massive star
forming clumps to be prime producers of massive super star clusters. Once
made these clusters will be the most likely to survive and 
observed at later times.

We thank the WFPC2 Investigation Definition Team, and the Space Telescope
Science Institute Archival Observer program for providing support
through the National Aeronautics and Space Administration for this work.

\end{document}